\documentclass{bmcart}

\usepackage{amsthm,amsmath}
\RequirePackage[authoryear]{natbib}
\RequirePackage{hyperref}
\usepackage[utf8]{inputenc} 

\usepackage{graphicx}
\usepackage{wrapfig}
\usepackage{epstopdf}

\startlocaldefs
\endlocaldefs

\begin{document}

\begin{frontmatter}

\begin{fmbox}
\dochead{Review}


\title{Toward epidemic thresholds on temporal networks: a review and open questions}


\author[
   addressref={vt-fiw},            
   corref={vt-fiw},                
   email={jack.leitch@vt.edu}      
]{\inits{JL} \fnm{Jack} \snm{Leitch}}
\author[
   addressref={vt-fiw,caracal},
   email={kathyalx@vt.edu}
]{\inits{KA} \fnm{Kathleen A.} \snm{Alexander}}
\author[
   addressref={vt-stat},
   email={sengupta@vt.edu}
]{\inits{SS} \fnm{Srijan} \snm{Sengupta}}


\address[id=vt-fiw]{                          
  \orgdiv{Department of Fish and Wildlife Conservation},
  \orgname{Virginia Tech},
  \cny{USA}
}
\address[id=caracal]{
  \orgname{CARACAL, Centre for Conservation of African Resources: Animals, Communities, and Land Use}, 
  \city{Kasane},
  \cny{Botswana}
}
\address[id=vt-stat]{%
  \orgdiv{Department of Statistics},
  \orgname{Virginia Tech},
  \cny{USA}
}


\end{fmbox}


\begin{abstractbox}

\begin{abstract} 
Epidemiological contact network models have emerged as an important tool in understanding and predicting spread of infectious disease, due to their capacity to engage individual heterogeneity that may underlie essential dynamics of a particular host-pathogen system. Just as fundamental are the changes that real-world contact networks undergo over time, both independently of and in response to pathogen spreading. These dynamics play a central role in determining whether a disease will die out or become epidemic within a population, known as the epidemic threshold. In this paper, we provide an overview of methods to predict the epidemic threshold for temporal contact network models, and discuss areas that remain unexplored. 
\end{abstract}


\begin{keyword}
\kwd{temporal networks}
\kwd{epidemiology}
\kwd{epidemic threshold}
\end{keyword}


\end{abstractbox}
%

\end{frontmatter}



\section*{Introduction}

Significant advances have been made in the mathematical modeling of infectious diseases since the seminal work of \cite{Kermack1927AEpidemics}. The ability to track, measure, predict, and control the spread of contagion has vital implications for the health and well-being of humans, domestic animals, and wildlife. However, there are still many challenges to the creation of models that are able to capture real-world dynamics in complex systems. In this paper, we aim to provide a survey of the important recent developments in infectious disease modeling, specifically temporal contact networks, and identify areas of future research. In particular, we aim to provide an exposition that can be accessible to network scientists and enhance their interest in this topic. We first present well-known epidemic thresholds for temporal networks under the approach of time-scale separation. Next, we introduce four foundational classes of temporal contact network models, with their associated characterizations of the epidemic threshold. We then provide an overview of factors that may influence the epidemic threshold, reviewing work done to date to incorporate these factors into the foundational models. Finally, we present a number of open questions related to computation of epidemic thresholds on temporal networks with broader connections to epidemiology and network theory. 

Modeling of infectious disease has taken many different forms, most notably  mean-field compartmental models, first introduced by \citet{Kermack1927AEpidemics}. Here, individuals are broken up into compartments where a characteristic is shared, as for example, disease state (e.g., susceptible (S), infected (I), recovered (R)), reflecting the infection dynamics of a particular host-pathogen system \citep{Keeling2007ModelingAnimals}. While compartmental models are computationally simple, theoretically tractable, and relatively easy to fit to observational data, they can be limited by their fundamental assumptions. Among these is the assumption of homogeneity in mixing, meaning that every individual has equal probability of interacting with every other individual. In reality, however, interactions between individuals vary widely in both number and intensity, and this heterogeneity can significantly impact disease spread \citep{Meyers2005, Rocha2011, Galvani2005DimensionsSuperspreading, Woolhouse1997HeterogeneitiesPrograms}. Approaches that allow the transmission function to change based on geospatial or environmental factors can add substantial flexibility and robustness to compartmental models by modifying the strict assumptions on homogeneity \citep{levin1996individuals,graham2004spatial,mccallum2001should,paull2012superspreaders,engering2013pathogen,white1996host}.

In contrast, agent-based models provide maximum flexibility in modeling heterogeneity of individual behavior and potential disease-transmitting contacts with others through stochastic simulation of individuals and their interactions \citep{Alexander2012ModelingCaveats}.   Simulations of contagion spread through the resulting synthetic population can then be used to predict disease dynamics. For example, in \cite{moore2009extending}, the authors demonstrated that by using agent-based models, ethno-epidemiological data can be integrated in a study of psychostimulant use, which is difficult to accomplish by using compartmental models. However, these models can be computationally intensive for large populations and require incorporation of extensive and detailed knowledge of behavior. Additionally, the fundamental structure and principles of agent-based models preclude analytical treatment, which can be a barrier to intuitive understanding of their dynamics (for a full review, see \cite{willem2017lessons}).

In recent years, epidemiological contact network models have emerged as an alternative that engages fundamental heterogeneity of behavior while offering potential for analytical tractability \citep{keeling2005implications}. Notable examples where the use of contact networks have led to improvements in prediction or understanding of infectious diseases include \cite{bengtsson2015using} and \cite{kramer2016spatial}. In these models each node represents an individual and each edge connecting a pair of nodes represents contact with potential for pathogen transmission. Each individual is assigned a disease state depending on the particulars of the host-pathogen system under study (e.g., SIR, SIS). We can then imagine a spreading process occurring on the network where contagion moves from infected nodes to non-infected nodes. The network structure also makes computation less expensive than agent-based models since the network as a whole can be modeled, instead of tracking each individual agent. 

\subsection*{Temporal Networks}
In real epidemiological contact networks, however, infectious disease propagation proceeds with concomitant changes in the underlying network \citep{Bansal2010}. Here, contact patterns may change seasonally, or in response to transient conditions that may prompt bursts of contacts \citep{Enright2018, Barabasi2005TheDynamics, Hamede2009ContactDisease, Bharti2011ExplainingImagery, Hill2016TransmissionCycle}. Furthermore, individuals may change their contacts over time either to avoid infection or in direct response to infection \citep{White2018CovariationOutcomes, Croft2011EffectReticulata, Hawley2011DoesDynamics, Verelst2016, Schaller2011TheSociality}. Indeed, a growing body of evidence suggests that pathogens can even modify individual host behavior directly in ways that are beneficial to the pathogen \citep{Berdoy2000FatalGondii, Goodman2011DiseasePlan}. The assumption that a contact network does not change significantly over time can result in profound mischaracterization of disease spread across the network \citep{Fefferman2007}. Temporal networks engage these dynamics, providing an opportunity to couple disease spreading and network evolution.

The distinctive characteristic of temporal contact networks is the ordering of contacts over time. Here, pathogen transmission progresses across a temporal network only through {\em time-respecting paths} \citep{Holme2012}, which are sequences of edges that follow one after another in time. A time-respecting path could act as a conduit for disease transmission from the starting node through the ending node, whereas a non-time-respecting path could not and is a dead-end for pathogen transmission. When stochastic transitions take place in the network (e.g., the removal of a node), the set of time-respecting paths get adjusted accordingly (e.g., the removed node cannot be a part of time-respecting paths in the future). Indeed, the proportion of paths that are not time-respecting can be considered a rough measure of the importance of modeling disease propagation using a temporal network rather than a static network \citep{Enright2018}.

The spread of pathogens through a connected population depends on the frequency and strength of interactions between the members of the population. Real-world contact networks are known to exhibit various temporal dynamics, such as concentrated bursts followed by long periods of inaction \citep{Barabasi2005TheDynamics} and periodic patterns due to seasonality \citep{Enright2018}. Such changes substantially affect the spread on infectious diseases in a population \citep{Zino2018}. Temporal contact networks, therefore, provide a valuable tool for studying the spread of infectious diseases. 

\subsection*{Epidemic Thresholds}

Particular interest has been given to the question of whether a pathogen, when introduced into an entirely uninfected population, will die out or lead to an epidemic. The outcome is dictated by the parameters of the system with respect to a condition known as the {\em epidemic threshold}. More than just a theoretical component, growing evidence provides that these thresholds exist in real-world host-pathogen systems \citep{Dallas2018ExperimentalThreshold}. The epidemic threshold is commonly specified in terms of the {\em basic reproduction number}, $R_0$, defined as the average number of individuals infected by a typical infected individual in an otherwise uninfected population \citep{Diekmann2000MathematicalInterpretation}. When $R_0>1$, the contagion may invade; when $R_0<1$, it dies out. More generally, the epidemic threshold can be expressed in terms of the critical value of one or more model parameters. Above the epidemic threshold, the pathogen invades and infects a finite fraction of the population. Below the epidemic threshold, the prevalence (total number of infected individuals) remains infinitesimally small in the limit of large networks \citep{Pastor-Satorras2015EpidemicNetworks}.

Many interventions are developed and implemented based on estimates of this value, seeking either to raise the threshold for the population (e.g., through targeted vaccination) \citep{shulgin1998pulse,wallinga2005measles} or to adjust the value of the critical parameter below the threshold (e.g., by reducing probability of transmission of a respiratory pathogen through use of face masks) \citep{pourbohloul2005modeling,Meyers2005}.

Recent years have seen significant advances in estimation of the epidemic thresholds for temporal contact networks. Throughout this paper we characterize epidemic threshold conditions using parameters of each particular epidemiological contact network model. Typically these will be stated in terms of the critical transmissibility $\lambda_c=\beta/\mu$ of the pathogen, where $\beta$ is the infection rate per effective contact and $\mu$ is the recovery rate, so that an epidemic is likely to occur when $\lambda>\lambda_c$. (See Table 1 for notation definition.)

\begin{table}[ht!]
  \caption{Common Symbols}
  \begin{tabular}{|p{2cm}|p{9.85cm}|}
    \hline
    
    Symbol & Definition and Description \\
    
    \hline
    
    $\mathbf{A}, \mathbf{B}, \ldots$ & matrices (bold upper case) \\
    
    $\mathbf{I}$ & the $n \times n$ identity matrix \\
    
    $\rho(\mathbf{A})$ & spectral radius (largest eigenvalue) of the matrix $\mathbf{A}$ \\
    
    $\left<\cdot\right>$ & statistical expectation \\
    
    $k_i$ & degree of the node $i$ of the network \\
    
    $k$ & the degree distribution of the network \\
    
    $\left<k\right>, \left<k^2\right>$ & first and second moments of the degree distribution of the network \\
    
    \hline
    
    $S(t), I(t), R(t)$ & number of susceptible ($S$), infected ($I$), and recovered/removed ($R$) individuals in the population at time $t$ \\
    
    $\beta$ & infection rate: probability of transmission of a pathogen from an infected individual to a susceptible individual per effective contact (e.g. contact per unit time in continuous-time models, or per time step in discrete-time models) \\
    
    $\mu$ & recovery rate: probability that an infected individual will recover per unit time (in continuous-time models) or per time step (in discrete-time models) \\
    
    $\lambda$ & transmissibility: the infection rate rescaled by $\mu^{-1}$ so that $\lambda = \beta/\mu$ \\
    
    $\lambda_c$, $\beta_c$, etc. & critical transmissibility, critical infection rate, etc., that define the epidemic threshold \\
    
    $\omega$ & rewiring rate or mixing rate \\
    
    $a$ & activity rate distribution for activity-driven networks \\
    
    $m$ & number of connections to other nodes formed by an active node at each time step, for activity-driven networks \\
    
    \hline
  \end{tabular}
  \label{tbl:notation}
\end{table}

\section*{Time Scale Separation}

Many papers studying disease spread on temporal networks have relied on time-scale separation techniques, an approach which assumes that the rate of spread of a pathogen across the network and the evolution of the network itself take place on distinct time scales. When the spreading process occurs much more rapidly than changes to the network, known as the {\em quenched regime}, the network can be fully characterized by a static adjacency matrix \citep{Wang2003EpidemicViewpoint}. In the opposite case, the {\em annealed regime}, the network changes are assumed to occur much faster than the disease spreading process \citep{Newman2002}. Annealed networks can then be well represented by an average adjacency matrix $\mathbf{\overline{A}}_{k_i,k_j}$, expressing the probability that two vertices of degree $k_i$ and $k_j$ are connected in the original network. We defer to \cite{Pastor-Satorras2015EpidemicNetworks} for a thorough review of the topic, but present a few key findings here for comparison with results discussed in later sections.

Under degree-based mean-field (DBMF) theory, the epidemic threshold for uncorrelated annealed networks can be estimated for both SIS dynamics \citep{Pastor-Satorras2001a} and SIR dynamics \citep{Moreno2002EpidemicNetworks} by
\begin{equation}
    \lambda_c = \frac{\left<k\right>}{\left<k^2\right>},
    \label{eqn:crit-trans-dbmf-an}
\end{equation}
where $\left<k\right>$ and $\left<k^2\right>$ are the first and second moments of the degree distribution of the network, respectively. This expression elegantly relates the critical transmissibility of a pathogen to the moments of the degree distribution, such that it is proportional to the average connectivity of the network and inversely proportional to fluctuations in the connectivity. Because annealed networks are characterized by their average adjacency matrix, the epidemic threshold described in Eq. (\ref{eqn:crit-trans-dbmf-an}) can be considered exact for annealed networks.

DBMF theory can be improved by accounting for the inability for a node to infect the node that infected it (as that node is assumed to be no longer susceptible). \citet{Newman2002} calculated the epidemic threshold in this case as
\begin{equation}
    \lambda_c = \frac{\left<k\right>}{\left<k^2\right>-\left<k\right>}.
    \label{eqn:crit-trans-dbmf-qn}
\end{equation}

For a comprehensive review of DBMF theory on networks with degree correlations, see \cite{boguna2003corr}. Other notable contributions in this area include those of \cite{eguiluz2002epidemic} and \cite{serrano2006percolation}, who derived the epidemic threshold for scale-free networks and clustered networks, respectively. In \cite{colizza2007invasion}, the authors derived the invasion threshold for heterogeneous metapopulation networks.

A key assumption of DBMF is that all nodes with the same degree are considered to be statistically equivalent. Individual-based mean-field (IBMF) theory, on the other hand, considers the individual probability that each node will be in a particular disease state at time $t$, thus it can only be applied to quenched networks. The IBMF approach is valid under the mean field assumption of independence between nodes' infectious states. In \cite{gomez2010discrete}, the authors proposed a discrete-time formulation of the problem to resolve a family of models that range from the so-called contact process to the so-called reactive process. Under this theory, the exact epidemic threshold for an arbitrary static network is given by  
\begin{equation}
    \lambda_c = \frac{1}{\rho[\mathbf{A}]},
    \label{eqn:crit-trans-ibmf}
\end{equation}
where $\rho[\mathbf{A}]$ is the spectral radius of the adjacency matrix $\mathbf{A}$. \cite{Boguna2002EpidemicNetworks}, \cite{Wang2003EpidemicViewpoint}, and \cite{Chakrabarti2008} derived this epidemic threshold for SIS dynamics. In \cite{van2009virus}, the authors established this epidemic threshold result under the $N$-intertwined Markov chain model. \cite{Prakash2010} later established that this estimate of the epidemic threshold holds for any disease propagation model and any static network topology. In \cite{wang2016predicting} and \cite{wang2017unification} the authors generalized this result to a broader class of networks. This result can be seen as consistent with the epidemic thresholds described in Eq. (\ref{eqn:crit-trans-dbmf-an}) and (\ref{eqn:crit-trans-dbmf-qn}). For example, under SIS dynamics, the annealed network equivalent to a given uncorrelated quenched network will have an average adjacency matrix with leading eigenvalue $\left<k^2\right>/\left<k\right>$ \citep{Castellano2010}.

Accurate modeling of real-world contact networks depends on availability of sufficiently detailed contact data, which can be difficult to obtain \citep{Eames2015SixModelling}. Quenched and annealed networks may still be useful in these cases, as they may allow capture of critical features of the network without requiring full contact information. For example, data that permits estimation of the fraction of time that contact occurs between each pair of individuals can inform a quenched network with edges weighted according to contact time \citep[e.g.,][]{Moslonka-Lefebvre2012WeightingNetworks, Stehle2011SimulationAttendees, Corner2003Social-networkVulpecula}. Though this approach neglects the ordering of contacts, it may provide actionable insights.

\section*{Foundational Temporal Network Models}

Over time, a few temporal contact network models have become established in the literature as a foundation on which the study of epidemic thresholds on temporal networks has been constructed. Here we introduce these models along with their associated characterizations of the epidemic threshold.

\subsection*{Neighbor-Exchange Model}
\citet{Volz2009} examined the effect of social mixing on SIR disease spreading for a family of temporal networks, providing an important early result that the computational characterization of the epidemic threshold depends not only on disease parameters of the system and static network topology, but also the rate at which the network changes over time. In this model, nodes are assigned a fixed random node degree according to a configuration model \citep{Molloy1998}. Then neighbor exchanges, in which pairs of edges are selected uniformly randomly and instantly swapped, occur as a Poisson process at a fixed mixing rate, $\omega$. Thus each individual maintains a fixed number of concurrent contacts while the identities of the contacts change stochastically over time.

Considering the probability generating function $g(k)$ for the degree distribution, the critical transmission rate is given by
\begin{equation}
    \beta_c = \frac{\mu\left(\mu+\omega\right)g'(1)}
                   {\left(\mu+\omega\right)g''(1) + \left(\omega-\mu\right)g'(1)},
\end{equation} 
and the critical mixing rate by
\begin{equation}
    \omega_c = \frac{\mu\left(\beta+\mu\right)g'(1) - \beta\mu g''(1)}
                    {\beta g''(1) + \left(\beta-\mu\right)g'(1)}.
    \label{eqn:ne_crit_mixing_rate}
\end{equation}
The relation for the critical mixing rate in Eq. (\ref{eqn:ne_crit_mixing_rate}) establishes two special conditions for the mixing rate: When $\lambda < \lambda_\textrm{lb} = g'(1)/\left(g''(1)+g'(1)\right) = \left<k\right>/\left<k^2\right>$, an epidemic cannot occur even when the mixing rate is arbitrary large. Similarly, when $\lambda > \lambda_\textrm{ub} = g'(1)/\left(g''(1)-g'(1)\right) = \left<k\right>/\left(\left<k^2\right>-2\left<k\right>\right)$, an epidemic may occur even when the mixing rate is zero (defining a static network). Between these two bounds, an epidemic may occur but only when $\omega > \omega_c$.

\subsection*{Activity-Driven Networks}

\citet{Perra2012} introduced the activity-driven network (ADN) model, which permits evolution of the network in discrete time through a flexible Markov process. Each node is assigned a time invariant activity rate $a_i$, according to a given probability distribution $a$. At time step $t$, each node becomes active with probability $a_i\Delta t$ and forms $m$ connections with $m$ other nodes selected uniformly at random. Disease transmission is then evaluated over all connections, and all connections are cleared prior to the next time step. The epidemic threshold for the activity-driven network with SIS dynamics is
\begin{equation}
    \beta_c = \frac{\mu}{m\left(\left<a\right>+\sqrt{\left<a^2\right>}\right)}.
    \label{eqn:adn-epidemic-threshold}
\end{equation}
Whereas the epidemic threshold in models discussed so far have been specified in terms of the degree distribution, here it computationally arises from the activity rate distribution and the connectivity parameter. This may not be surprising, given that, in a sense, they implicitly define the effective degree distribution of the network.

While not an activity-driven network per se, \citet{Taylor2012} developed a similar model involving random link activation and deletion with per-potential-link ($\alpha$) and per-link ($\omega$) rate parameters, respectively. In addition, they introduced a constraint on link activation in terms of the maximum allowable per-node degree ($M$), finding that this parameter alone could influence the outcome of an epidemic under SIS dynamics for any particular pair of rate parameters $\alpha$ and $\omega$. This suggests the potential for controlling epidemics by simply limiting the total number of contacts per individual. Their model allows for analytic calculation of $R_0$, and \citeauthor{Taylor2012} were among the first to establish the limited value of $R_0$ in predicting epidemic outcome in temporal networks.

Further work has been done on activity-driven networks by \cite{Starnini2013}, who examined topological properties of time-aggregated activity-driven networks, \cite{Starnini2014}, who confirmed the epidemic threshold defined in Eq. (\ref{eqn:adn-epidemic-threshold}) from a temporal percolation perspective. Another key contribution was made by \cite{Zino2016} who presented a continuous-time formulation of the problem. In recent work, \cite{petri2018simplicial} and \cite{iacopini2019simplicial} have extended the activity driven model to higher-order interactions represented by so-called simplicial complexes.

\subsection*{Temporally-Switching Networks}

\cite{Prakash2010a} examined the epidemic threshold under SIS dynamics for arbitrary temporal networks, represented by a sequence of $T$ static network snapshots with adjacency matrices $\mathcal{A} = \left\{\mathbf{A_1}, \mathbf{A_2}, \ldots \mathbf{A_T}\right\}$. They show that the epidemic threshold is then characterized by the condition
\begin{equation}
    \rho[\mathbf{M}] = 1, \qquad \mathbf{M} = \prod_{t=1}^{T}\left(\left(1-\mu\right)\mathbf{I} + \beta \mathbf{A_t}\right),
    \label{eqn:tsn_prakash_threshold}
\end{equation}
where $\rho[\mathbf{M}]$ is the spectral radius of the system-matrix $\mathbf{M}$. When $T=1$, we recover the epidemic threshold condition (Eq. \ref{eqn:crit-trans-ibmf}) estimated by \cite{Chakrabarti2008} and others for static networks. \citeauthor{Prakash2010a} note the generality of their approach, as the period $T$ can be arbitrary, and snapshots may be repeated as desired within the sequence $\mathcal{A}$.

\cite{Valdano2015a, Valdano2015} also derived an estimate of the epidemic threshold for temporally-switching networks, obtaining a similar outcome using a different method. In their model, temporally-aggregated snapshots of arbitrary temporal networks are mapped to a multilayer network defined by directed edges representing potential transmission of a pathogen between individuals from one time step to the next. Reducing the tensor representation of this multilayer network to matrix form and considering the Markov process for transition from one time step to the next under SIS and SIR dynamics, they derive the epidemic threshold condition
\begin{equation}
    \rho[\mathbf{P}]^{1/T}=1,\qquad \mathbf{P}=\prod_{t=1}^{T}\left(\left(1-\mu\right)\mathbf{I}+\beta \mathbf{A_{t}}\right),
\end{equation}
which is mathematically equivalent to Eq. (\ref{eqn:tsn_prakash_threshold}). The matrix $\mathbf{P}$ can be interpreted as encoding in each entry $P_{ij}$ the probability that node $j$ is infected at time $t=T$, and that this infection originated at node $i$ at time $t=0$ and traveled along one or more time-respecting paths (which is valid around the disease-free state and under the assumption of non-interaction among paths). \citeauthor{Valdano2015a} therefore term the matrix $\mathbf{P}$ as {\em infection propagator}. 

The common approach of \citeauthor{Prakash2010a} and \citeauthor{Valdano2015}, of using a sequence of temporally-aggregated network snapshots, appears to be highly flexible, as it imposes no assumptions on the structure of the temporal network, and the estimated epidemic threshold fully incorporates the topological and temporal dynamics of the network. However, the approach of \citeauthor{Valdano2015} assumes a period boundary condition, $\mathbf{A_{(T+t)}} \equiv \mathbf{A_t}$, where the temporal network snapshots are repeated over time. The authors note that this may result in time-respecting paths that are not present in the original temporal network, but whether this affects the estimated epidemic threshold is an open question. In \cite{Valdano2015}, the authors present numerical results indicating that it is possible to determine an optimal minimum period length that will allow for full characterization of the epidemic threshold.

\cite{Valdano2018} later extended their approach to continuous-time temporal networks and found that in weakly-commuting networks (where the adjacency matrix at time $t$ commutes with the aggregated adjacency matrix up to that time), the epidemic threshold is the same as if it were computed on the time-averaged adjacency matrix $\mathbf{\overline{A}}$, as in Eq. (\ref{eqn:crit-trans-ibmf}).
\cite{speidel2016temporal} studied the epidemic threshold of the SIS model on arbitrary temporal networks and established its connection to the commutator norm. \cite{Speidel2017} assert that the epidemic threshold decreases as the time step between network snapshots increases, and is bounded above by the epidemic threshold for the continuous-time model. They also examine \textit{commutation} \citep{speidel2016temporal} of the adjacency matrices, finding that, when all pairs of snapshot adjacency matrices commute, the epidemic threshold is the same as that of a continuous-time temporal network.

\section*{Exploring Key Factors}

Various authors have presented extensions to the base temporal contact network models outlined above, accounting for factors that are believed to affect epidemic dynamics. Broadly, these include characteristics and dynamics of the underlying contact networks, infectious disease dynamics, and the complex interactions that can arise between them, all potentially influencing the epidemic threshold.

\subsection*{Social Structure}

Epidemiological contact networks are naturally driven by contacts between individuals and, whether the population under study is human, domestic animals, or wildlife, social structure largely dictates the nature of these contacts. In wildlife, it is believed that social structure has evolved in part to protect populations from spread of infectious disease \citep{Rozins2018SocialMammal, Sah2017UnravelingNetworks}. In humans as well, widespread social and cultural norms often serve as a barrier to disease spreading \citep{Schaller2011TheSociality, Fincher2012Parasite-stressReligiosity}. On the other hand, it is believed that pathogens have evolved in some cases to exploit social structure in order to further their survival. The effect of social structure on disease spreading has been studied extensively for static contact networks \citep{Salathe2010DynamicsStructure, Huang2007EpidemicStructure, Wu2008HowNetworks, Stegehuis2016EpidemicStructures, Liu2005EpidemicNetworks}, and inclusion of social structure in temporal contact networks may provide crucial insights into real-world systems.

\cite{Nadini2018} recently explored the effect of explicit static community structure on disease spreading in activity-driven networks. In their model, each node is randomly assigned to a single community at time $t=0$, with community sizes taken from a heavy-tailed distribution. When a node is activated, rather than forming connections with other nodes selected uniformly at random from the entire population, connections are formed with nodes selected uniformly at random from the same community or social group with probability $\mu$ and from a different community or social group with probability $1-\mu$. In this way, strength of the social structure is parameterized by $\mu$. The relation defining the epidemic threshold for these modular activity-driven networks has no closed-form solution, but some conclusions can be drawn from its form. As $\mu \rightarrow 0$, the modular structure of the network vanishes, and the threshold for both SIS and SIR dynamics is the same as in the activity-driven network without modularity \citep{Perra2012}. As $\mu \rightarrow 1$, however, strong modularity induces a difference between the epidemic threshold for SIR and SIS dynamics. When modularity is strong, the effects appear to be driven by infected individuals having an increased probability of contact with members of their own social group or community. For SIR dynamics, these repeated contacts are increasingly with recovered individuals, in which case contagion spread will be reduced, inhibiting pathogen transmission and movement beyond the social group. For  SIS  dynamics, the  movement of individuals from infected  to susceptible status  allows sustained transmission and the potential for pathogen endemicity depending on group size and pathogen extinction potential. Pathogen persistence in a social group provides greater probability of extra-group transmission and spread.

Though this model incorporates social structure and changing contacts over time, we must also recognize that social structure itself evolves, and, depending on the time scale on which changes occur relative to disease dynamics, may significantly inhibit or promote disease spreading. For example, with the population growth and recovery of the South American sea lion (\textit{Otaria flavescens}), significant changes in social composition and spatial distribution of colonies has been observed \citep{grandi2008social}, modifications that will have important influence on pathogen transmission and persistence dynamics. Pathogen infection itself can also influence movement behaviors of individuals feeding back to modify dispersal behavior and social structure across the population, as observed in banded mongoose (\textit{Mungos mungo}) infected with the novel tuberculosis pathogen \textit{Mycobacterium mungi} \citep{fairbanks2014impact}.

\subsection*{Temporal Contact Patterns}

Fundamental to temporal contact networks is the concept that each individual's social contact patterns can change over time. \cite{Holme2015} examined the effects of these changes on spreading processes, finding that, for a set of empirical networks, exact times and order of contacts is less predictive of disease outbreaks than the beginning and end of contact and the overall intensity of contact during that period.

Real social networks include both {\em strong ties}, contacts that are made repeatedly, frequently, or for long periods, as well as {\em weak ties}, which are isolated, sporadic, or of short duration. Incorporating both types of contact patterns can be challenging, as strong ties typically require models to be non-Markovian, which tends to result in reduced analytic tractability. 

\cite{Sun2015} studied the effects of strong and weak ties on disease spreading using activity-driven network models with memory. In their non-Markovian model, an active node with memory of $n_i$ previously-contacted nodes will contact a new node with probability $1/(n_i+1)$ and a previously-contacted node with probability $n_i/(n_i+1)$. Through numerical simulation, they found that for SIR dynamics the memory effect increased the epidemic threshold, while for SIS dynamics the memory effect lowered the epidemic threshold. This result was also reported by \citet{Karsai2015}, and is consistent with the effects of community structure on epidemic spreading, as described by \cite{Nadini2018}.

Most temporal contact network models assume that contacts occur as a Poisson process, but it has been observed that human behavior tends to result in events occurring in concentrated bursts alternating with long periods of inaction. This phenomenon is known as {\em burstiness} \citep{Barabasi2005TheDynamics}.

\cite{Zino2018} examined the effect of burstiness in contact patterns on disease spreading in temporal networks. To do this, they altered the activity-driven network model by replacing the standard Poisson activation process for each node with a Hawkes process \citep{Hawkes1971SpectraProcesses}. Four time invariant parameters were assigned to control to each node's Hawkes process: (i) the jump $J_i>0$, which defines the strength of the self-excitement effect, (ii) the forgetting rate $\gamma_i>0$, which defines how quickly excitement is forgotten, (iii) an initial activity rate $a_i(0)>0$, and (iv) a background activity rate, $\hat{a}_i>0$. It is easily seen that if, $\forall i$, $J_i=0$ and $a_i(0)=\hat{a}_i$, the result is the standard activity-driven network model with Poisson activation process. In the interest of analytic tractability, \citeauthor{Zino2018} assume the same Hawkes process for all nodes, therefore assigning the same jump and forgetting rate to each node. Under these assumptions, when $J<\gamma$, the epidemic threshold is given by
\begin{equation}    
    \lambda_c = \frac{1-\frac{J}{\gamma}}
                     {\left<\hat{a}\right>+\sqrt{\left<\hat{a}^2\right> + \frac{J^2}{2\gamma}\left<\hat{a}\right>}}
    \label{eqn:adn-hawkes-epidemic-threshold}
\end{equation}
When $J=0$, Equation \ref{eqn:adn-hawkes-epidemic-threshold} reduces to the epidemic threshold for the SIS model for standard activity-driven networks \citep{Perra2012}. The above epidemic threshold is less than the epidemic threshold for the standard activity-driven network. Estimates of the epidemic threshold for this model using Monte Carlo numerical simulations find that the epidemic threshold is also reduced under SIR dynamics. The key result here is that models that do not account for burstiness may significantly overestimate the epidemic threshold.

In real-world social interactions, it is possible to observe coordinated bursts of contacts. For example, wildlife may gather at a common water source during a draught, temporarily placing them in close contact. To date, the effects of this phenomenon remain unexplored.

Contacts may also exhibit periodic patterns, e.g., due to seasonal changes in behavior and social interaction \citep{Enright2018}. The effect this has on disease spreading has not been explicitly examined in temporal networks, though most temporal network models appear to be capable of extension by simply repeating the time period. Indeed, this may be a strength of approaches such as the infection propagator of \cite{Valdano2015}, which assumes periodic boundary conditions.

\subsection*{Node Set Changes}

Temporal network models typically assume a closed population, but in most populations demographic changes such as births, deaths, emigration, and immigation are likely to occur on a time scale commensurate with the disease spreading process. Indeed, it has long been known from compartmental models that demographic changes can significantly affect epidemic dynamics \citep{Keeling2007ModelingAnimals}.

\cite{Guerra2012} and \cite{Demirel2017} investigated the interaction between continuous network growth and disease spreading. Network growth occurred through preferential attachment \citep{Barabasi1999EmergenceNetworks}, a process known to result in the creation of scale-free networks. Scale-free topologies are known to have no epidemic threshold in the thermodynamic limit \citep{Pastor-Satorras2001a}.  A real example of this phenomenon was studied in \cite{rushmore2014network}, where it was found that highly central individuals in primate social contact networks also tend to be larger-bodied individuals who just happen to encounter more pathogens. However, under SIR dynamics, nodes having high degree as a result of preferential attachment are also more likely to become infected and then removed (assuming the rate of recovery is fast enough), thus raising the epidemic threshold.

\subsection*{Disease Dynamics}

While significant attention has been given to how network topology and dynamics affect disease spread, comparatively little progress has been made in understanding the impacts of individual heterogeneity on disease dynamics. Individuals can vary in their relative susceptibility, infectivity, latency, and/or duration of the infectious period in real-world populations. Furthermore, these individual parameters may change over time, as these parameters might be associated with aging, changes in reproductive status, or in response to medical treatment. Coinfection can also introduce additional heterogeneity where one pathogen may induce partial immunity \citep{dietz1979epidemiologic} or, alternatively, increased susceptibility to another pathogen infecting the same host.



\cite{Darbon2019} examined the importance of accounting for variation in infection duration, reduction or extension of which could result in fewer or greater secondary infections, respectively. They calculated the epidemic threshold for three real-world networks using the infection propagator approach of \citep{Valdano2015}, concluding that failing to account for this type of heterogeneity could result in significant mis-estimation of the epidemic threshold. 




\subsection*{Behavioral Response to Pathogen Spreading: Adaptive Networks}

It is well known that contact patterns may change over time in response to disease spreading. Humans and animals are known to avoid contact with infected individuals or reduce their interactions overall in response to awareness of disease, a phenomenon known as {\em social distancing} or {\em protective sequestration} \citep{Reluga2010GameEpidemic}. Indeed, quarantine or reducing exposure to a pathogen by reducing contact with others is well established as a public health intervention. Individuals may also have reason to increase their interactions in order to achieve exposure to a pathogen, thus promoting disease spread, sometimes in the interest of preventing future infection \citep{henry2005,lopes2016infection,ezenwa2016host,cole2006complexity,aleman2009accounting}. Individual behavior can also be influenced by the clinical response to pathogen infection. For example, symptoms from infection such as lethargy may temporally reduce an individual's contacts with others. Pathogens themselves may increase interactions in infected individuals in ways that promote their spread \citep{poulin2010parasite, Lefevre2009InvasionInteractions}.


\citet{Gross2006} examined epidemic dynamics on discrete-time adaptive networks under SIS disease dynamics. In their model, at each time step, susceptible nodes disconnect from each adjacent infected node with probability $\omega$ (the re-wiring rate) and form a new link with another susceptible node selected uniformly at random. Thus individuals protectively modify their contact patterns in response to local knowledge of the disease spreading process. This rewiring process introduces spreading dynamics that are not found in static networks, including bistability characterized by two thresholds: the {\em persistence threshold}, which is the minimum transmissibility for an already-established disease to remain endemic, and the {\em invasion threshold}, which is equivalent to the epidemic threshold that we examine herein \citep{Marceau2010}. Under this adaptive rewiring behavior, \citeauthor{Gross2006} find that the epidemic threshold is characterized by
\begin{equation}
    \beta_c = \frac{\omega}{\left<k\right>\left[1-\exp{(-\omega/\mu)}\right]}.
\end{equation}

\cite{Risau-Gusman2009} consider a model based on \citet{Gross2006}, but rewiring at each time step with a node selected uniformly at random from the entire population, irrespective of the target node's disease status. Under this arguably more realistic model, the epidemic threshold is characterized by
\begin{equation}
    \beta_c = \frac{\omega + \mu}{\mu\left<k\right> - \mu}.
\end{equation}

One key dynamic observed by \citeauthor{Gross2006} is that, while rewiring acts as a barrier to a disease becoming epidemic, over time it also induces formation of closely-connected communities comprised solely of susceptible individuals. Because they are more densely connected, these communities therefore have a lower average epidemic threshold than the entire network. The model of \citeauthor{Risau-Gusman2009} does not exhibit this effect, as new connections are not formed preferentially with susceptible individuals, but rather without regard to infection status.
We note that \cite{shaw2008fluctuating} made an important early contribution to this area by studying the fluctuating dynamics of the SIRS model on adaptive networks.

\cite{Rizzo2014} investigated the effects of decreased activity rate in infected individuals in activity-driven networks. They define activity rate multipliers $\eta_S$ and $\eta_I$ for susceptible and infected individuals respectively, where $\eta_I<\eta_S$, then found the epidemic threshold to be given by
\begin{equation}
    \lambda_c=\frac{2}{m\left(\left(\eta_S+\eta_I\right)\left<x\right> + \sqrt{\left(\eta_S-\eta_I\right)^2\left<x\right>^2+4\eta_S \eta_I\left<x^2\right>}\right)}.
\end{equation}
Thus activity reduction for infected individuals increases the epidemic threshold. When $\eta_S=\eta_I=\eta$, there is no difference in activity rates between susceptible and infected individuals, and the epidemic threshold matches that found by \cite{Perra2012} (Eq. \ref{eqn:adn-epidemic-threshold}). 

\cite{Kotnis2013} used activity-driven networks to investigate the effects of social distancing in response to global awareness of infectious disease spreading. Their model considers two base activity rates, $a_h$ for healthy individuals and $a_\textrm{inf}$ for infecteds, where $a_h\geq a_\textrm{inf}$. Thus infected individuals are assumed to have a potentially lower activity rate as a result of either being aware of their infected state or due to clinical symptoms of infection. Susceptible individuals are assigned an activity rate $a_\textrm{sus}(I)=a_h e^{-\delta\cdot I(t)}$, incorporating a risk perception factor $\delta$ that causes activity rates to decrease as the number of infecteds $I(t)$ increases. The resulting epidemic threshold for the SIS model is given by
\begin{equation}
    \lambda_c = \frac{1}{m \left(a_h e^{-\delta I} + a_\textrm{inf}\right)}.
\end{equation}
\cite{Rizzo2014} used a similar activity-driven network model to demonstrate that the epidemic threshold increases when susceptibles reduce their activity rate in response to global awareness of infection.

In reality, individuals may not be informed of the global prevalence of infection within the population, and may only be aware of infecteds within their local neighborhood in the network. \cite{Hu2018} considered this case for activity-driven networks using an SAIS model, including an Alert state to represent individual awareness of risk and therefore preventative behavior, based on the number of infected and alert neighbors. Alert individuals adopt a preventative behavior for $h$ time steps, and then return to their normal behavior. This duration $h$ can be used to represent temporary preventative behavior such as wearing masks ($h=1$) or a more permanent intervention such as vaccination ($h=\infty$). The authors find that when the duration of the preventative behavior is short, risk awareness has no effect on the epidemic threshold, but longer durations serve to increase the epidemic threshold.

\cite{Moinet2018} examined the effects of awareness in activity-driven networks with memory. To account for adoption of preventative behavior due to awareness as a result of contact with infected individuals, the probability of infection spreading from an infected individual to a susceptible individual $i$ is specified by $\beta_i(t)=\beta \exp{[-\delta n_I(i)_{\Delta T}]}$, where $\delta$ represents the strength of the awareness and $n_I(i)_{\Delta T}$ is the proportion of contacts with infected individuals within the time interval $\left[t-\Delta T,t\right]$. They report that the epidemic threshold for both SIS and SIR dynamics is unaffected by awareness in activity-driven networks without memory, but awareness increases the epidemic threshold in activity-driven networks with memory. It should be noted, though, that this effect is only seen in finite networks and does not appear to be present in the limit of large networks.


\subsection*{Multiple Modes of Transmission}

Some pathogens may have more than one method of transmission. For example, human immunodeficiency virus (HIV) may be transmitted horizontally through sexual contact or blood transfer or vertically during pregnancy or childbirth. These modes of transmission are the result of substantially different types of contact between individuals, with each having a different associated probability of transmission \citep{Patel2014EstimatingRisk}. Additionally, the temporal concurrency or ordering of these different types of contacts may influence the spread of disease through the population as a whole \citep{Morris2010TimingPrevalence}.

\section*{Open Questions}

Both theory and epidemiological applications of temporal networks are active areas of continuing research. Despite the importance of explicitly modeling change in networks over time, temporal networks are still much less popular in disease modeling than static methods, as the theory and methodology of temporal networks are actively developing areas of study \citep{Pellis2015}.

For classical epidemiological models and most static contact networks, the basic reproduction number, $R_0$, is directly related to the spreading extent and therefore the epidemic threshold. \cite{Holme2015R0} observed that this is not necessarily the case for temporal networks. Examining correlations between $R_0$ and the spreading extent on a variety of empirical human contact networks, they found that temporal and topological characteristics of temporal networks had different effects on the relationship between $R_0$ and epidemic spreading.

In nearly all cases, static and temporal contact network models are composed of closed and isolated populations, but in our globally connected world human contact networks commonly have have far-reaching connections. This may be addressed using a combination of contact network models at different population scales. What dynamics emerge when network changes occur over time at these different scales? And, when we require closed models for analytic or computational tractability, can we account statistically for assumed interactions with individuals external to the network?

As typically defined, the epidemic threshold characterizes the conditions (i.e., the parameters of the epidemic model) under which an epidemic is likely to occur. A related problem of interest is to estimate the probability of an epidemic given a particular set of conditions. This could be of substantial practical importance in scenarios where the pathogen transmission rate and the infection rate are known (or can be estimated within a reasonable margin of error), and the problem of interest is to determine the chances of an epidemic in a given population. The probability of epidemic is not typically estimated in conjunction with the epidemic threshold. Perhaps this could be accomplished by framing the epidemic threshold as a function of the likelihood of epidemic rather than as a singular value. This could provide a more nuanced way to measure cost and benefit of interventions.

\subsection*{Random Graph Models and Community Detection}

Most of the existing literature on epidemic thresholds considers the network to be deterministic and fully observed.
This approach does not allow for uncertainty quantification or future predictions of interactions.
In statistical network analysis, there has been a great deal of emphasis on random graph models that can be used to address this issue.
Well-studied random graph models include stochastic blockmodels and its variants \citep{holland1983stochastic,karrer2011stochastic,senguptapabm}, latent space models \citep{hoff2002latent,handcock2007model}, and exponential random graph models \citep{snijders2006new,robins2007introduction}, to name a few.
Well studied statistical models for time-varying networks include dynamic latent space models \citep{sewell2015dyn,sewell2016weighted,sewell2015analysis}, dynamic stochastic blockmodels \citep{matias2017statistical}, and temporal exponential random graph models (TERGM) \citep{krivitsky2014separable}.

Further, the social structure approach of \cite{Nadini2018} explores the situation where different communities have different recovery rates, under the assumption that the community structure of the network is known.
In many real-world situations, this information may not be available.
In such cases it would be useful to identify the communities by using community detection techniques like spectral clustering \citep{rohe2011spectral,qin2013regularized,sengupta2015spectral,lei2015consistency} or likelihood modularity maximization \citep{bickel2009nonparametric, zhao2012consistency}.
Connecting the existing results on epidemic thresholds to statistical network models and community detection techniques can greatly broaden the applicability of these results.

Understanding community structure can lead to improved efficacy of targeted immunization strategies. Typical community detection approaches assign individuals to a single community, but in real-world networks for social species individuals are often members of more than one community. Recently \cite{Ghalmane2019} and \cite{Ghalmane2019} have addressed this issue for static networks by introducing a centrality measure that accounts for overlapping community structure (see \cite{Cherifi2019} for a discussion of application to targeted immunization). While potentially challenging due to changing community structure over time, extending this approach to temporal networks could provide a method for further optimizing selective immunization and other targeted interventions.

\subsection*{Local Structures and Network Dynamics}
Local structures, i.e., interaction patterns localized to a small part of the network, can play a significant role in the transmission of a diseases through the network.
For example, the presence of unusually dense subgraphs or anomalous cliques (fully connected subgraphs) in a network can accelerate the spread of the disease, while on the other hand, chain-like structures can decelerate the spread of the disease.
There has been a rich thread of the network science literature focused on detection of such structures \citep{alon1998finding,dekel2014finding,feige2010finding,butucea2013detection,verzelen2015community,arias2014community,sengupta2018anomaly,komolafe2017statistical,miller2015spectral,miller2010toward,MillerSparsePCA}.
Connecting these detection techniques to epidemic thresholds and understanding the impact of local structures on disease dynamics is an important open question.

Network dynamics is another aspect of network structure that has significant influence over the spread of a disease.
In particular, changes in network dynamics can have significant impact on disease transmission. 
In \cite{sengupta2018discussion}, the authors pointed out that network monitoring can refer to two cases, where the network is deterministic with some disease or information  being propagated through it, and  where the network is itself changes over time and is considered to be a time series.
The first problem is called \textit{fixed network surveillance}  and a lot of work has been done particularly on computer networks \citep{jeske2018statistical}.
In this framework, the network itself is fixed over time, and the intensity of information or disease propagation through each edge is considered to be randomly evolving over time.
The second problem is called \textit{random network surveillance} and is a rapidly emerging topic in network science \citep{priebe2005scan,yu2018detecting,woodall2016,wilson2016modeling,zhao2018aggregation,zhao2018performance}.
In this framework, the network itself changes over time following some time series model.
Both problems have important connections with disease transmission and epidemic thresholds, and studying these connections is still an open question.

\subsection*{Connecting Theory to Practice}

While significant advances have been made in the theory of estimating epidemic thresholds on temporal networks, there has been comparatively little progress in understanding how to apply this theory. One key application is that of risk assessment in case of real-world infectious diseases and outbreaks. To date, this has largely been an unexplored area of research. One notable recent contribution in this direction was made by \cite{darbon2018network}, where the authors linked the epidemic threshold to the prevalence of Bovine brucellosis in Italy.

In most of the current literature on epidemic thresholds, the underlying contact network is assumed to be fully known and observed. In practice, however, this is very unlikely to be the case in general. A fully observed sequence of contact networks might be available for small human groups or animal colonies, but it is not feasible or practical to obtain the detailed sequence of contact networks for, say, all residents of New York City. In absence of complete knowledge on how exactly the network changes over time, the epidemic threshold formulae would not allow prediction. In such circumstances, an alternative is to collect small partial samples from the network, and estimate network features (e.g., average degree, largest eigenvalue) related to epidemic thresholds from such samples. The problem of \textit{network sampling} has been well-studied in the network science community \citep{dasgupta2014estimating,feige2006sums,gjoka2010walking,goldreich2008approximating}. Compared to current literature where network features are assumed to be fixed, estimated values from network sampling are going to be random, leading to uncertainty in epidemic threshold computation. \cite{Genois2015} consider some approaches to this problem but note their significant limitations. 

On the other hand, in certain epidemiological circumstances (e.g., pathogens with an environmental reservoir or aquatic pathogens at smaller scales) it might not be necessary to obtain the full network dynamics in order to make predictions. Rather, a simpler model based on the important parts of the data might be sufficient. Understanding what kind of data is needed in which epidemiological context is an important direction of future research. 
\newline

\noindent As temporal network theory advances, so too will our ability to analyze spreading processes on these networks. Considerable progress has already been made toward estimating epidemic thresholds on temporal networks, but our toolbox still needs to be developed and refined further. We would encourage epidemiologists, disease ecologists, and network scientists to work on these open problems. Further research toward addressing these unknowns will allow us to capture essential biological and ecological factors, thus moving us toward our goal of improved tracking, prediction, and intervention in the spread of infectious disease.


\begin{backmatter}

\section*{Abbreviations}
ADN:~activity-driven network;
DBMF:~degree-based mean-field;
HIV:~human immunodeficiency virus;
IBMF:~individual-based mean-field;
SAIS:~susceptible-aware-infected-susceptible;
SI:~susceptible-infected;
SIR:~susceptible-infected-recovered/removed;
SIRS:~susceptible-infected-recovered-susceptible;
SIS:~susceptible-infected-susceptible;
TERGM:~temporal exponential random graph model

\section*{Acknowledgements}
  Not applicable.

\section*{Funding}
  Salary support for JL was provided by the National Science Foundation (Grant Number 1518663) (KA) as part of the joint NSF-NIH-USDA Ecology and Evolution of Infectious Diseases program.

\section*{Availability of Data and Materials}

\section*{Author's contributions}
  JL contributed a review of techniques for estimating epidemic thresholds in static and temporal networks. KA provided connections to modeling concepts in disease ecology and epidemiology. SS provided connections to relevant approaches in network theory. All authors read and approved the final manuscript.

\section*{Competing interests}
  The authors declare that they have no competing interests.


\bibliographystyle{bmc-mathphys} 
\nocite{settings}
\bibliography{references}        

\end{backmatter}

\end{document}